\documentclass[twocolumn]{aastex631}
\usepackage{color}
\usepackage[titletoc]{appendix}
\usepackage{amsmath}
\usepackage{amssymb}
\usepackage{mathtools}
\usepackage{upgreek}
\usepackage{comment}
\usepackage{enumitem}
\usepackage{natbib}
\usepackage{graphicx}
\usepackage{bm}
\usepackage{totcount}
\usepackage{multirow}
\usepackage{url}
\usepackage{tabularx}
\usepackage[T1]{fontenc}

\newcommand{\T}{{\rm T}}

\newcommand{\etastar}{\eta_{\star}}
\newcommand{\etaGR}{\eta_{\rm GR}}

\defcitealias{2020ApJ...902L...5P}{P20}
\defcitealias{2013A&A...553A..39C}{C13}

\shortauthors{Louden \& Millholland}
\shorttitle{Polar Neptunes are Stable to Tides}

\begin{document} 

\title{Polar Neptunes are Stable to Tides}

\author[0000-0003-3179-5320]{Emma M. Louden}
\affiliation{Department of Astronomy, Yale University, New Haven, CT 06511, USA}
\email{emma.louden@yale.edu}

\author[0000-0003-3130-2282]{Sarah C. Millholland}
\affiliation{Department of Physics, Massachusetts Institute of Technology, Cambridge, MA 02139, USA}
\affiliation{MIT Kavli Institute for Astrophysics and Space Research, Massachusetts Institute of Technology, Cambridge, MA 02139, USA}

\begin{abstract}

There is an intriguing and growing population of Neptune-sized planets with stellar obliquities near $\sim90^{\circ}$. One previously proposed formation pathway is a disk-driven resonance, which can take place at the end stages of planet formation in a system containing an inner Neptune, outer cold Jupiter, and protoplanetary disk. This mechanism occurs within the first $\sim10$ Myr, but most of the polar Neptunes we see today are $\sim$Gyrs old. Up until now, there has not been an extensive analysis of whether the polar orbits are stable over $\sim$Gyr timescales. Tidal realignment mechanisms are known to operate in other systems, and if they are active here, this would cause theoretical tension with a primordial misalignment story. In this paper, we explore the effects of tidal evolution on the disk-driven resonance theory. We use both $N$-body and secular simulations to study tidal effects on both the initial resonant encounter and long-term evolution. We find that the polar orbits are remarkably stable on $\sim$Gyr timescales. Inclination damping does not occur for the polar cases, although we do identify sub-polar cases where it is important. We consider two case study polar Neptunes, WASP-107 b and HAT-P-11 b, and study them in the context of this theory, finding consistency with present-day properties if their tidal quality factors are $Q \gtrsim 10^4$ and $Q \gtrsim 10^5$, respectively.

\end{abstract}

\section{Introduction}
\label{sec: Introduction}
One of the many unexpected discoveries within exoplanetary science is that a planet's orbital angular momentum vector need not be aligned with its host star's spin vector. The angle between these two vectors is called the ``stellar obliquity'' or stellar spin-orbit misalignment. The solar obliquity is $7^{\circ}$ when measured using the mean plane of the planets \citep{2005ApJ...621L.153B}. This relatively small value is consistent with expectations from planet formation. However, for exoplanets, a broad range of stellar obliquities have been observed, and they exhibit various trends with stellar and planetary properties. (See \citealt{2022PASP..134h2001A} for a review.) For instance, hot Jupiters orbiting hot stars have high obliquities \citep{2010ApJ...718L.145W}, as do super-Earths and sub-Neptunes orbiting hot stars \citep{2021AJ....161...68L, 2024ApJ...968L...2L}.

The so-called ``polar planets'' are a class of systems with $\sim90^{\circ}$ stellar obliquities. \cite{2021ApJ...916L...1A} computed the 3-dimensional stellar obliquity ($\Psi$) for 57 systems by combining constraints on stellar inclination angles with Rossiter-McLaughlin measurements of sky-projected stellar obliquities ($\lambda$). The findings revealed a significant number of systems with nearly perpendicular orbits ($\Psi \approx 80^{\circ} - 120^{\circ}$), indicating a statistical preference for polar orbits over a full range of obliquities when using frequentist tests. Recent works using Bayesian methods have conflicting evidence for an overabundance of perpendicular orbits. \cite{Siegel_2023, 2023AJ....166..112D} do not find strong evidence for the polar planets whereas \cite{2023A&A...674A.120A} and \cite{knudstrup2024obliquitiesexoplanethoststars} find evidence that such a peak does exist, especially when restricting attention to the sample of sub-Saturn planets and hot Jupiters orbiting F stars.  
Firm conclusions on the statistical nature of the polar planet population await a larger sample size. Regardless, the known polar planet systems are still dynamically puzzling. 

\begin{figure*}
    \centering
    \includegraphics[width=0.95\textwidth]{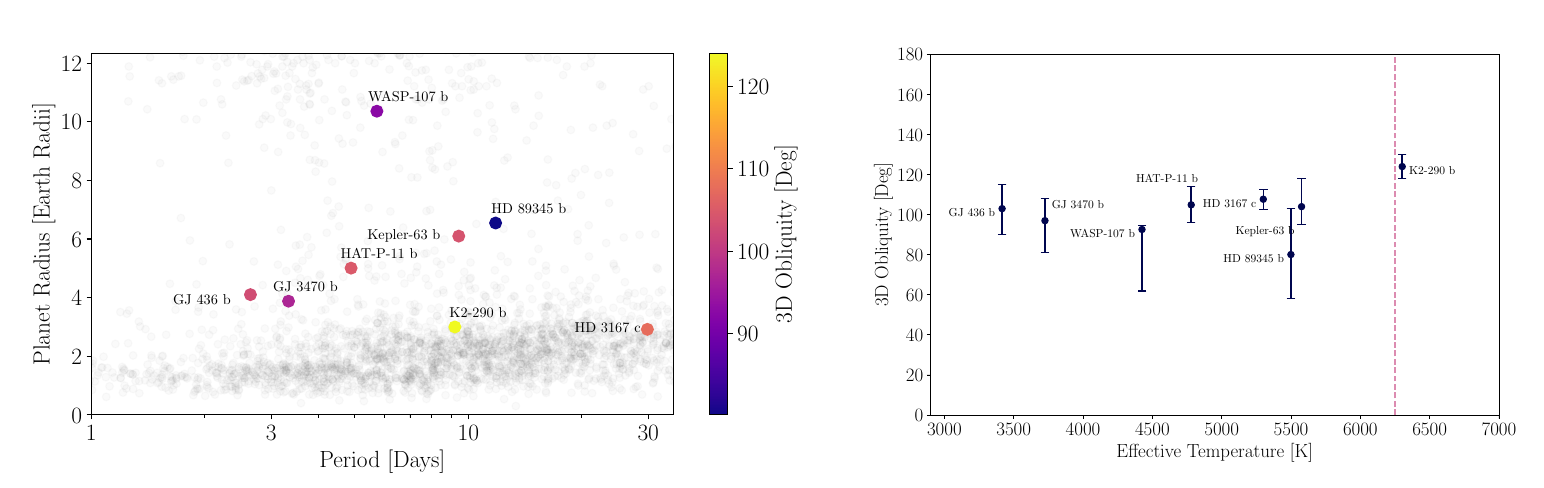}
    \caption{\textbf{The current population of observed polar Neptunes.} We define polar Neptunes as planets with radii in the range $R_p = 1.5 - 7 \ R_{\oplus}$ or masses in the range $M_p = 10 - 50 \ M_{\oplus}$ and measured 3D stellar obliquities in the range $\Psi = 80^{\circ} - 125^{\circ}$. The left panel shows this sample in planet radius/orbital period space, with the coloration according to the 3D stellar obliquity. The right panel shows their 3D stellar obliquities, $\Psi$, versus the stellar effective temperature. The data is compiled from TEPCat \citep{2011MNRAS.417.2166S} and originally from \cite{
    2010ApJ...710.1724B, 2011ApJ...743...61S, 2018AJ....155..255Y, 2017A&A...604A.110A, 2017AJ....153..205D, 2016ApJ...829L...9V, 2019A&A...631A..28D, 2021A&A...654A.152B, 2019MNRAS.484.3522H, 
    2022ApJ...931L..15S, 
    2022A&A...663A.160B, 2022MNRAS.514...77M, 2023A&A...669A..63B}. Note that there are several additional planets that have $|\lambda|$ values suggestive of polar orbits, but they don't yet have constrained $\Psi$ values. These planets include WASP-156 b \citep[$\lambda = {105.7^{+14.0}_{-14.4}}^{\circ}$,][]{2018A&A...610A..63D, 2023A&A...669A..63B}, HATS-38 b \citep[$\lambda = {-108^{+11}_{-16}}^{\circ}$,][]{2020AJ....160..222J, 2024arXiv240618631E}, and WASP-139 b \citep[$\lambda = {-85.6^{+7.7}_{-4.2}}^{\circ}$,][]{2017MNRAS.465.3693H, 2024arXiv240618631E}. }
    \label{fig:polar_population}
\end{figure*}

About half of the polar planets are  warm Neptune or super-Neptune-sized
planets. These polar Neptunes share a variety of distinguishing characteristics: moderately eccentric ($\sim0.05 < e \lesssim 0.3$) and polar orbits, puffy atmospheres that show evidence for mass loss in several cases, and evidence for exterior massive companion planets in
some cases (see Figure \ref{fig:polar_population}). Examples include WASP-107 b \citep{2017A&A...604A.110A, 2017AJ....153..205D, 2024arXiv240600187Y}, HAT-P-11 b \citep{2010ApJ...710.1724B, 2011ApJ...743...61S, 2018AJ....155..255Y, 2024arXiv240519510A, 2024arXiv240519511L}, HD 3167 c \citep{2016ApJ...829L...9V, 2019A&A...631A..28D, 2021A&A...654A.152B}, and K2-290 b \citep{2019MNRAS.484.3522H}. Moreover, most polar Neptunes reside in or near the hot Neptune desert \citep{2016A&A...589A..75M, 2023A&A...669A..63B, 2023A&A...674A.120A}, a sparsely-populated region of period-radius space that is thought to be sculpted by photoevaporation \citep{2012ApJ...761...59L, 2013ApJ...775..105O, 2014ApJ...795...65J} and potentially high-eccentricity migration \citep{2018MNRAS.479.5012O}. 

The mechanism(s) responsible for the unusual orbits of the polar Neptunes should also explain these other features. However, there is not yet consensus as to the origins of the polar Neptunes. Several theoretical scenarios have been proposed to explain 3D obliquities near $\Psi \approx 90^{\circ}$. 
\cite{Lai_2012} showed that tidal dissipation within the host star can strand a system at a $90^{\circ}$ obliquity. This mechanism works through inertial wave dissipation in stars with convective zones, but it probably doesn't explain all perpendicular systems. 
In addition, primordial magnetic interactions between the protostar and the protoplanetary disk can torque the disk to high inclinations, with the planets then forming in these misaligned states \citep{2011MNRAS.412.2799F, 2011MNRAS.412.2790L}.
Similarly, \cite{2024MNRAS.tmpL..55C} recently proposed that the central star in a polar planet system may have started as tight binary system that subsequently merged, and the polar planet formed in a protoplanetary disk that was perpendicular to the original binary orbital plane. As for another possibility, von Zeipel-Kozai-Lidov cycles and planet-planet scattering have long been invoked to explain spin-orbit misalignments \citep[e.g.][]{2007ApJ...665..754F, 2008ApJ...678..498N, 2011ApJ...742...72N,
2012ApJ...751..119B, 2016MNRAS.456.3671A}, but they have gained extra attention recently in the context of the polar planets \citep[e.g.][]{2023ApJ...943L..13V, 2024arXiv240600187Y, 2024arXiv240519511L}.

Separate from these ideas, another compelling theory for generating $\sim90^{\circ}$ stellar obliquities is disk-driven resonance, which was proposed by \cite{2020ApJ...902L...5P} (hereafter \citetalias{2020ApJ...902L...5P}). This mechanism is envisioned to operate within the first $\sim 10$ Myr in a system containing an inner Neptune-sized planet, outer giant planet, and a transition disk. As the disk dissipates, a secular inclination resonance is encountered between the inner Neptune and distant giant planet, which tilts the Neptune's orbit up to $90^{\circ}$. This theory accounts for the high obliquities, the non-zero eccentricities, and the Jovian companions with large mutual inclinations sometimes found in the polar Neptune systems. Its ability to explain all of these features at once makes disk-driven resonance particularly compelling. This theory is our primary focus in this paper. 

\begin{figure*}
    \centering
    \includegraphics[width=\textwidth]{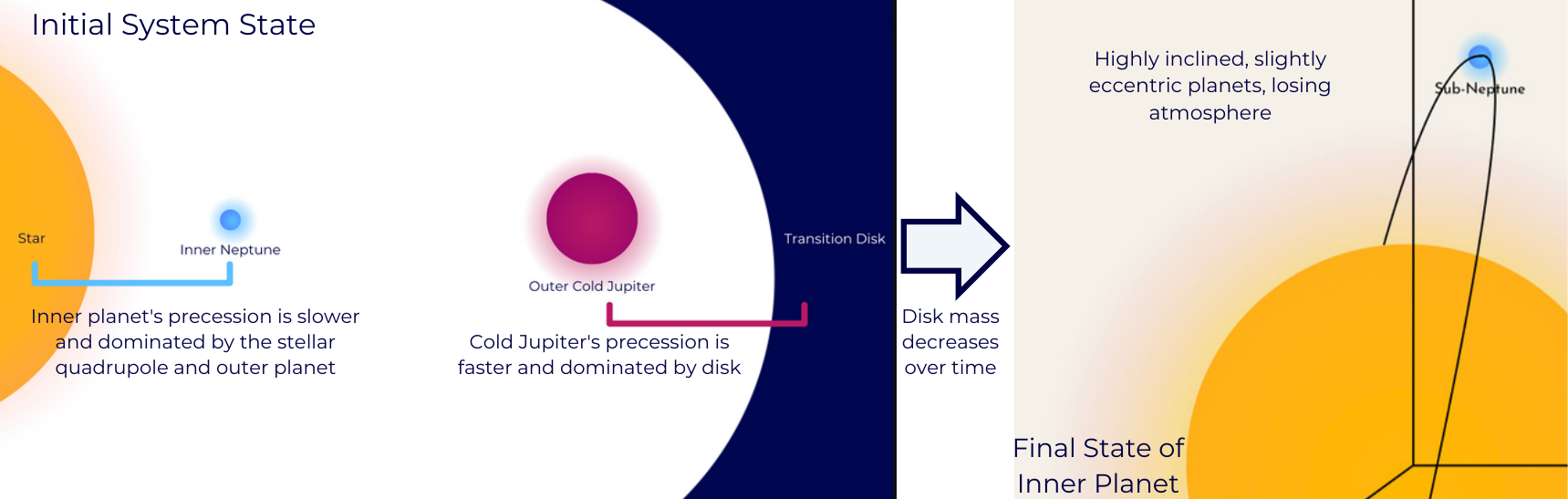}
    \caption{\textbf{Schematic of the disk-driven resonance theory for the origins of polar Neptunes.} The starting configuration (left) consists of a star with an approximately coplanar inner Neptune, outer cold Jupiter, and transition disk. The nodal precession of the outer planet is initially faster than that of the inner planet due to the presence of the disk, but as the disk mass decays, eventually a secular inclination resonance is encountered where the nodal precession rates of the two planets are equal. The inner Neptune's inclination then increases, as does the eccentricity in some cases. The inner Neptune is left in a highly misaligned state that is often polar (right).}
    \label{fig:syssetup}
\end{figure*}

A key question is whether disk-driven resonance creates polar orbits that are long-term stable. Disk-driven resonance occurs at the very early epochs of formation, but the polar Neptunes we see today are mostly billions of years old. Within those $\sim$Gyrs of time, the orbits can be modulated by effects like tidal and stellar evolution, and these effects can potentially be significant. For example, tidally-driven orbital realignment is thought to operate for hot Jupiters orbiting cool stars, explaining the differences in the hot star and cool star populations \citep[e.g.][]{2010ApJ...718L.145W, 2012MNRAS.423..486L, 2014ApJ...784...66X, 2014ApJ...790L..31D, 2022ApJ...927...22S, 2023A&A...674A.120A, 2024ApJ...967L..29Z}. There can also be more subtle effects, like inclination damping driven by interactions between the planet's spin and orbit \citep{2013A&A...553A..39C}. There has not yet been a thorough investigation of tidal effects in the context of the disk-driven resonance mechanism. It is crucial to determine whether there is any impact on the mechanism's viability and the polar planet's long-term stability.



In this paper, we explore the degree to which the disk-driven resonance theory for polar Neptunes is influenced by tidal evolution. We investigate both short-term and long-term effects on the orbital and spin evolution. The organization of the paper is as follows. Section \ref{sec:deepdive} describes the disk-driven resonance mechanism and the properties of a system that affect its outcome. Section \ref{sec: Early evolution} focuses on the system's early evolution while the disk is still present. We explore how tides impact the initial excitation into a polar orbit. Next, in Section \ref{sec:Long term}, we study the long-term evolution of the system. We specifically explore whether tidal effects may erase the initial polar orbit. Finally, we discuss the results in Section \ref{sec:Conclusions} and explore two case studies of observed near-polar Neptunes, HAT-P-11 b and WASP-107 b.

\section{Formation of Slightly Eccentric Polar Neptunes through Disk-Driven Resonance}
\label{sec:deepdive}

\citetalias{2020ApJ...902L...5P} introduced the disk-driven resonance theory to explain polar Neptunes. Here a close-in ($\lesssim 0.1$ AU) Neptune-sized planet interacts with both an exterior ($\sim1-5$ AU) giant planet and an outer, slowly-depleting protoplanetary disk of gas and dust with the inner $\sim 5$ AU evacuated. This is motivated by the existence of observed transition disks with inner regions depleted of gas and dust \citep{2014prpl.conf..497E}. Initially, perturbations to the outer planet's orbit are dominated by the disk, while the inner planet's orbit is perturbed primarily by the outer planet and the star's quadrupolar gravitational field (oblateness). The inner planet is assumed to have zero eccentricity at the start (see Figure \ref{fig:syssetup}).  

The dynamical transition to a polar orbit begins with the establishment of secular inclination resonance. The disk induces nodal precession of the outer planet's orbit, $\dot{\Omega}_{\mathrm{out}}$, at a rate proportional to the disk mass, which we assume decreases smoothly and slowly. The outer planet, in turn, induces nodal precession of the inner planet's orbit, $\dot{\Omega}_{\mathrm{in}}$.  Initially, $|\dot{\Omega}_{\mathrm{out}}| > |\dot{\Omega}_{\mathrm{in}}|$. As the transition disk dissipates, $|\dot{\Omega}_{\mathrm{out}}|$ decreases until the precession rates of the two planets match and
a scanning secular resonance \citep{1980Icar...41...76H, 1981Icar...47..234W} is crossed. The inclination of the
inner planet is forced to increase because it stays in resonance by maintaining the same nodal precession frequency as the outer planet, and the precession frequency slows as the inclination rises. 


The inner planet's inclination growth stops at a critical inclination $I_{\mathrm{crit}}$, which, under certain conditions, is polar. This is a generalization of the critical inclination in the von Zeipel-Kozai-Lidov mechanism \citep{2015ApJ...799...27P, 2015MNRAS.447..747L, 2016ARA&A..54..441N}. That is, as soon as the critical inclination is reached, the inner planet undergoes eccentricity excitation that detunes the secular resonance and halts the inclination growth. If $I_{\mathrm{crit}}$ is near $\sim90^{\circ}$ or undefined, then the planet can reach a polar orbit before the resonance is detuned. 

The critical inclination is dictated by the relative perturbations on the inner planet from the outer planet, general relativity (GR), and stellar oblateness. In the absence of GR and stellar oblateness, the critical inclination would be equal to $39.2^{\circ}$. However, the short-range forces suppress eccentricity excitation due to the periapse precession they induce \citep[e.g.][]{2015MNRAS.447..747L}, and thus a larger inclination is required to induce eccentricity excitation. The critical inclination can be calculated using the dimensionless quantities $\eta_{GR}$ and $\eta_{\star}$. Here $\eta_{GR}$ measures the strength of GR perturbations relative to the planet-planet interactions, and $\eta_{\star}$ similarly measures the strength of the stellar quadrupole perturbations relative to the planet-planet interactions. They are defined as 
\begin{equation}
    \eta_{\star} = \frac{2 J_2 M_{\star}}{m_{2}} \frac{R_{\star}^2 a_{2}^3}{a_{2}^5}(1-e_{2}^2)^{3/2}
\end{equation}
\begin{equation}
    \eta_{GR} = \frac{8 G M_{\star}}{c^2} \frac{a_{2}^3}{a_{1}^4} \frac{M_{\star}}{m_{2}}(1-e_{2}^2)^{3/2},
\end{equation}
where $M_{\star}$ and $R_{\star}$ are the stellar mass and radius, $c$ is the speed of light, $a_1$ is the semi-major axis of the inner Neptune, $a_2$ and $e_2$ are the semi-major axis and eccentricity of the outer cold Jupiter, and $m_2$ is its mass. In the equation for $\eta_{\star}$, $J_2$ is the coefficient of the quadrupole moment of the star's gravitational field, and it depends on the stellar rotation period $P_{\star}$, the Love number $k_{2\star}$ and $R_{\star}$ as \citep[e.g.][]{1939MNRAS..99..451S}
\begin{equation}
J_2 \approx \frac{k_{2\star}}{3}\frac{4\pi^2}{P_{\star}^2}\frac{R_{\star}^3}{G M_{\star}}.
\end{equation}

\begin{figure}
    \centering
\includegraphics[width=\columnwidth]{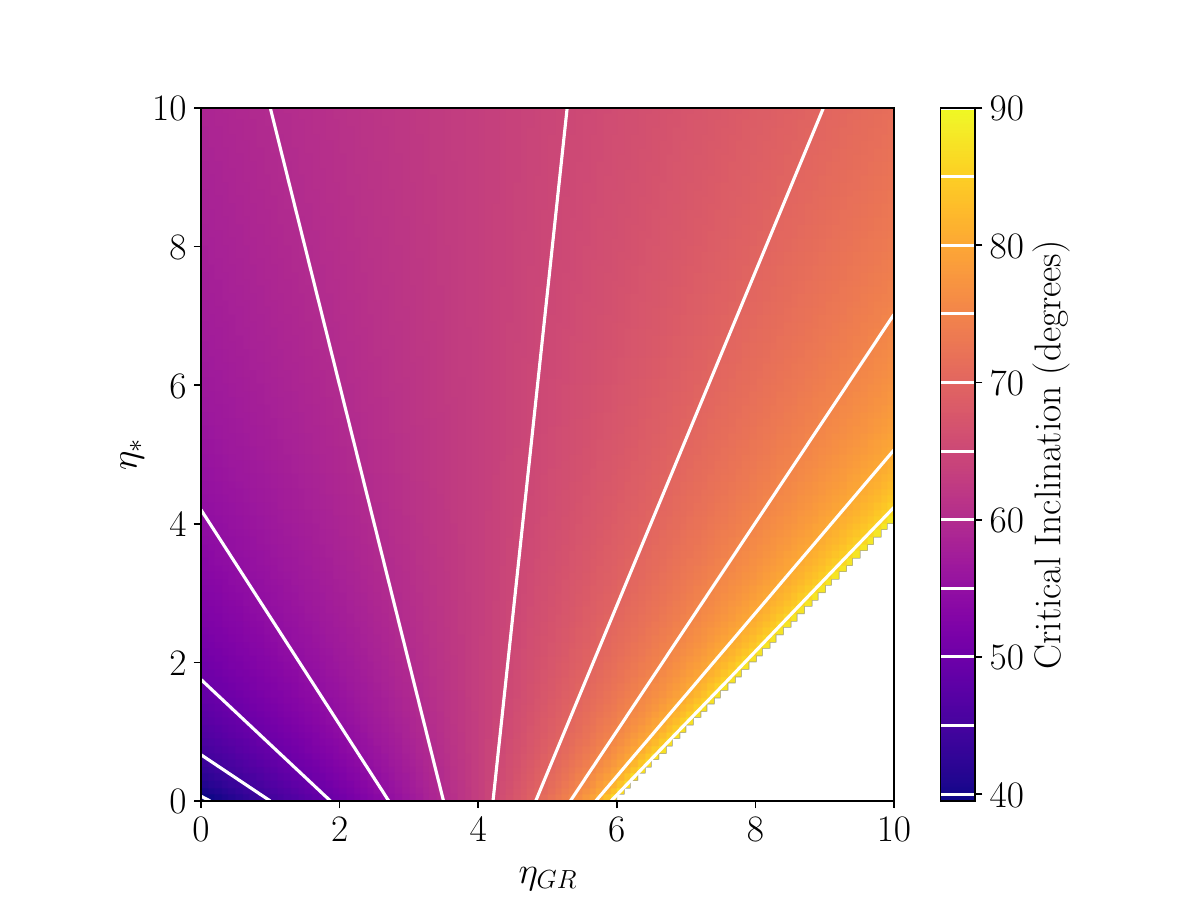}
    \caption{\textbf{Variation of the critical inclination with $\eta_{\star}$ and $\eta_{\mathrm{GR}}$.} The colorbar indicates the value of the critical inclination (equation \ref{eq: I_crit}), and contours are shown in $5^{\circ}$ increments. The white region indicates where $I_{\mathrm{crit}}$ is undefined (since the argument of the arcsin is greater than 1). Such cases would reach a polar orbit without undergoing eccentricity excitation.}
    \label{fig:TugOfWar}
\end{figure}

In terms of  $\eta_{\star}$ and $\eta_{\mathrm{GR}}$, the critical inclination is given by
\begin{equation}
    I_{\mathrm{crit}} = \sin^{-1}\left(\frac{4 + 4 \etastar + \etaGR}{ 10 + 5 \etastar}\right)^{1/2}.
\label{eq: I_crit}
\end{equation}
Figure \ref{fig:TugOfWar} shows how the combined values of $\eta_{\star}$ and $\eta_{\mathrm{GR}}$ determine the critical inclination. For $\etaGR \geq 6 + \etastar$, the argument of the arcsine is greater than 1 and thus all inclinations are stable, meaning the planet would reach a polar orbit. We also note that the behavior in the parameter space in Figure \ref{fig:TugOfWar} changes at $I_{\mathrm{crit}} = 63.4^{\circ}$, since apsidal precession from stellar oblateness is prograde for inclinations $< 63.4^{\circ}$ and retrograde for $> 63.4^{\circ}$. 

To summarize, the inner planet can reach a polar and possibly eccentric orbit, but the outcome is dependent upon a ``tug-of-war'' between the effects of the stellar oblateness, the GR effects, and the cold Jupiter. This competition is summarized succinctly in the quantities $\eta_{\star}$ and $\eta_{\mathrm{GR}}$, which are most sensitively dependent on six key parameters of the system: mass and radius of the star ($M_{\star}$, $R_{\star}$), rotation period of the star ($P_{\star}$), mass and semi-major axis of the cold Jupiter ($m_{2}$, $a_{2}$), and semi-major axis of the inner Neptune ($a_{1}$).

\section{Early Evolution With Tides}
\label{sec: Early evolution}

The disk-driven resonance theory could be affected by tidal evolution at multiple epochs in time. First, there is the early $\lesssim10$ Myr period during the disk phase when the resonance is encountered and the inclination (and sometimes eccentricity) excitation occurs. Second, after the polar orbit is established, there is the long $\sim$ Gyrs period during which orbital realignment might occur. We will investigate these phases separately, starting with the early evolution. Here we explore how the initial resonant encounter is impacted by tides. 

The initial resonance excitation involves a complicated set of interactions between the two planets, disk, and oblate star. We opt to perform direct $N$-body simulations to fully capture these interactions and the corresponding tidal and spin dynamics. Since such simulations are expensive, we only investigate a representative case study rather than assess the full parameter space. We use a numerical integrator that evolves the system using instantaneous accelerations in the framework of \cite{2002ApJ...573..829M}. This code self-consistently accounts for the accelerations from (1) standard Newtonian gravity, (2) the oblateness of the host star, (3) constant-time lag tides, and (4) the outer protoplanetary disk. It evolves the planetary orbits and spins simultaneously. The code was originally developed and described in \cite{2019NatAs...3..424M}, and further details can be found therein. The disk effects were added in \cite{2020AJ....160..105S}, but the disk was taken to be infinite. Here we model the disk as a finite transition disk between $R_{\mathrm{in}}$ and $R_{\mathrm{out}}$ with surface density $\Sigma = \Sigma_0 (r/R_{\mathrm{out}})^{-\alpha}$. The gravitational potential is given by \cite{2013ApJ...769...26C} as
\begin{equation}\label{eq:chen}
\begin{aligned}
\Phi &= -\frac{-\alpha + 2} {1 - \eta^{-\alpha + 2}} \frac{G M_{\text{disk}}}{R_{\text{out}}} \left[ \frac{1 - \eta^{1 - \alpha}}{1 - \alpha} \right. \\
& \left. + \frac{-1 + \eta^{-1 - \alpha}}{1+\alpha} \frac{r_p^2}{2 R_{\text{out}}^2} \left(-1 + \frac{3}{2} \sin^2\theta_p\right) \right],
\end{aligned}
\end{equation}
where $\eta = R_{\mathrm{in}}/R_{\mathrm{out}}$, $M_{\mathrm{disk}}$ is the mass of the disk, $r_p$ is the radial position of the planet, and $\theta_p$ is the polar angle of the planet. The disk mass decreases over time according to $M_{\mathrm{disk}} = M_{\mathrm{disk},0}/[1 + t/t_{\mathrm{disk}})]^{3/2}$.

We run two simulations of a representative synthetic system with all being equal except one simulation includes tides and the other does not. The parameters of the system are the same as the simulation in the rightmost panel of Figure 2 of \citetalias{2020ApJ...902L...5P}, so as to aid comparison to their results. This case corresponds to a resonant encounter that yields a nearly polar and eccentric planet. The system consists of a Solar-mass star, a Neptune-mass and Neptune-radius planet at 0.07 AU with an initial eccentricity of $0.01$ and inclination of $1^{\circ}$, a $4 \ M_{\mathrm{Jup}}$-mass planet at $2$ AU in a circular orbit with an inclination of $5^{\circ}$, and a transition disk with edges $R_{\mathrm{in}} = 3$ AU, $R_{\mathrm{out}} = 30$ AU and an initial mass of $50 \ M_{\mathrm{Jup}}$. The power law exponent of the disk surface density is set to $\alpha = 1$, and the disk decay timescale is set to $t_{\mathrm{disk}}=1$ Myr. The star has a radius of $1.3 \ R_{\odot}$, rotation period of 7 days, and Love number of $k_{2\star} = 0.2$. For the simulation of the planet that includes tides, we set the Love number of the planet to $k_{2p} = 0.4$ and tidal quality factor equal to $Q = 10^5$.

\begin{figure*}
    \centering
    \includegraphics[width=0.49\linewidth]{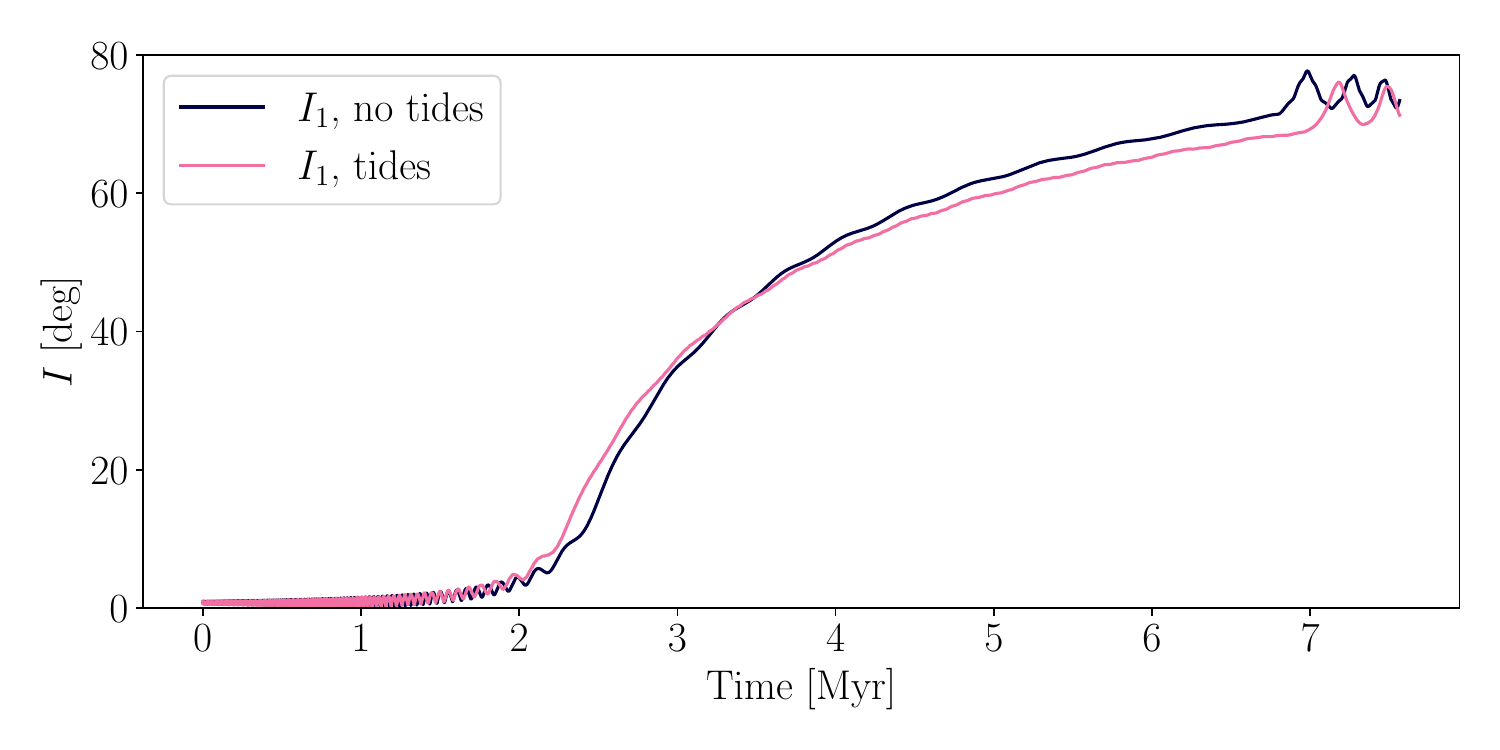}
    \includegraphics[width=0.49\linewidth]{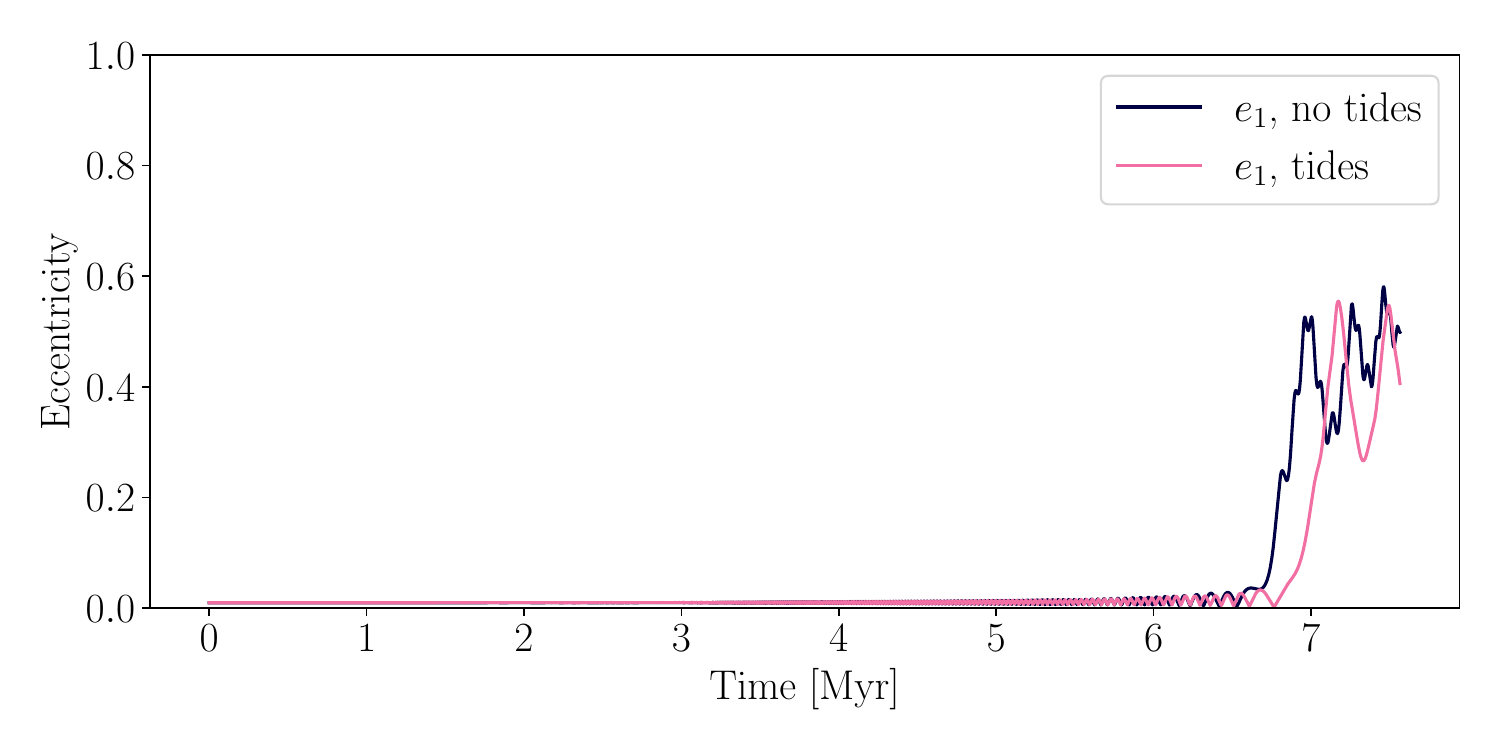}

    \caption{\textbf{Short-term $N$-body simulations with and without tides.} We show the early evolution of a synthetic system with representative parameters (see details in the text) as it encounters the disk-driven inclination resonance. The magenta and blue lines show the results without tides and with tides, respectively. The left panel show the inner and outer planets' inclinations, while the right panel show their eccentricities. Notably, the tides-free simulation replicates the result shown in  \citetalias{2020ApJ...902L...5P}, thus validating our numerical method. The simulation with tides shows a similar inclination evolution, suggesting that tides do not significantly affect the near-polar outcome. }
    \label{fig:ShortTermEvolution}
\end{figure*}

Figure \ref{fig:ShortTermEvolution} shows the results of the simulations. We show the inclination and eccentricity evolution of the inner Neptune-mass planet. 
Focusing first on the case without tides, we see that this set of parameters corresponds to a case in which the inner Neptune encounters the inclination resonance and tilts to $I_{\mathrm{crit}} \sim 70^{\circ}$, at which point the eccentricity grows rapidly and the resonance is detuned. The inclination growth is then stalled just above $70^{\circ}$. The behavior is the very similar to that seen in \citetalias{2020ApJ...902L...5P}, which simultaneously validates our code and stands as an encouraging sign since we are using a different numerical method ($N$-body, whereas \citetalias{2020ApJ...902L...5P} used secular simulations).


As for the simulation with tides turned on, the results are pretty similar overall. The inner planet's inclination evolution is nearly identical to the case without tides; it increases to the same near-polar critical inclination with a small lag. The eccentricity also increases when the critical inclination is reached, but the eccentricity growth is slightly delayed and muted by the tidal damping. We expect that further damping would be observed if the simulation was run for longer. For completeness, we note that we explored additional simulations with different system parameters (particularly the tidal quality factor of the inner planet) and found more complicated eccentricity behavior in some simulations. In these cases, the eccentricity decreased rapidly after its initial excitation and then stabilized at a lower value around $e\sim0.1$.
Exploring these dynamics across a broader parameter space is beyond the scope of this work. However, we can at least take away from this case study that tidal evolution does not significantly affect the inclination evolution but it can impact the eccentricity excitation following the resonant encounter. A more exhaustive parameter exploration could be considered in future work.

To round out our analysis of the early evolution, we also explore the behavior of the inner Neptune's spin vector during the disk-driven resonance encounter. Tidal dissipation in a planet is driven not only by its eccentricity but also its ``planetary obliquity'' (the angle between the planet's spin vector and the orbit normal vector) \citep{2010A&A...516A..64L}. Obliquity tides are theorized to drive orbital evolution and atmospheric inflation in a variety of exoplanetary systems \citep[e.g.][]{2019NatAs...3..424M}. It is thus relevant to investigate whether we expect an obliquity enhancement in the inner Neptune, which could affect the subsequent orbital evolution. Planetary obliquities can be excited by secular spin-orbit resonances, which occur when the planet's spin-axis precession frequency becomes commensurable with one of its orbit nodal precession frequencies. In Appendix \ref{sec: secular spin-orbit resonance}, we investigate this scenario in detail. We find that the parameter space in which the inner Neptune would obtain a high obliquity through a secular spin-orbit resonance does not overlap with the parameter space for it to obtain a polar orbit through disk-driven resonance. Therefore, we do not anticipate that the inner Neptune would experience significant obliquity tides, at least not due to obliquity excitation from a simultaneous resonance crossing.

\section{Long-Term Secular Evolution with Tides} \label{sec:Long term}

\begin{table*}
\centering
\caption{\textbf{Simulation variables and sampling scheme}}
\begin{tabular}{|l|l|l|l|}
\hline
Parameter & Description & Sampling distribution & Default value \\
\hline
$M_{\star}$ & Stellar mass (M$_{\odot}$) & Uniform(0.5, 3) & 1 \\
$m_1$ & Inner planet mass (M$_{\mathrm{Jup}}$) & Uniform(0.003, 0.09) & 0.05 \\
$m_2$ & Outer planet mass (M$_{\mathrm{Jup}}$) & Uniform(1, 10) & 4 \\
$a_1$ & Semi-major axis of inner planet (AU) & Uniform(0.02, 0.2) & 0.05 \\
$a_2$ & Semi-major axis of outer planet (AU) & Uniform(2, 5) & 2 \\
$e_1$ & Eccentricity of inner planet & Rayleigh(0.05) & 0.05 \\
$e_2$ & Eccentricity of outer planet & Uniform(0, 0.3) & 0 \\
$\omega_1$ & Argument of pericenter of inner planet (deg) & Uniform(0, 360) & Random \\ 
$\omega_2$ & Argument of pericenter of outer planet (deg) & Uniform(0, 360) & Random \\ 
$\Omega_1$ & Longitude of ascending node of inner planet & Fixed & 0 \\ 
$\Omega_2$ & Longitude of ascending node of outer planet & Fixed & $\pi$ \\ 
\hline
$P_{\star}$ & Stellar rotation period (days) & Uniform(10, 20) & 10 \\
$R_{\star}$ & Stellar radius (R$_{\odot}$) & Uniform(0.5, 1) & 1 \\
\hline
$R_1$ & Radius of inner planet (R$_{\odot}$) & Set by M-R Relation & N/A \\
$R_2$ & Radius of outer planet (R$_{\odot}$) & Set by M-R Relation & N/A \\
$k_{2\star}$ & Apsidal constant for star & Uniform(0.01, 0.1) & 0.01 \\
$k_{2,1}$ & Apsidal constant for inner planet & Uniform(0.2, 0.4) & 0.3 \\
$k_{2,2}$ & Apsidal constant for outer planet & Uniform(0.2, 0.4) & 0.3 \\
$t_{v,\star}$ & Viscous time for star & Uniform(1000, 5000) & 4000 \\
$Q_1$ & Tidal quality factor for inner planet & Uniform($10^3$, $10^7$) & N/A \\
$C_{\star}$ & Dimensionless moment of inertia of star & Uniform(0.05, 0.1) & 0.07 \\
$C_1$ & Dimensionless moment of inertia of inner planet & Uniform(0.15, 0.35) & 0.25 \\
$C_2$ & Dimensionless moment of inertia of outer planet & Uniform(0.15, 0.35) & 0.25 \\
\hline
\end{tabular}
\label{tab:random_values}
\end{table*}

With a clear picture of the impact of tides on the short-term initial phase of resonance capture, we now turn to the subsequent long-term $\sim$Gyrs period. The observed polar Neptunes are billions of years old and have been subject to tidal forces throughout their lifetime. 
Here we explore a wide range of parameters and model the long-term evolution to test if the tides cause a meaningful change like destabilization of the high inclination. Such an outcome would cast doubt on the theory that the polar orbits were established early on.

\subsection{Simulation set-up}

The $N$-body integrator we used in the last section is too slow to run over the long timescales of interest here. Instead, we perform secular (orbit-averaged) simulations using \texttt{KOZAIPY}\footnote{\url{https://github.com/djmunoz/kozaipy}}. \texttt{KOZAIPY} is a \texttt{Python} package for integrating the secular equations of motion of few-body, hierarchical systems including tidal friction. The equations are the same as in \cite{2007ApJ...669.1298F} and originally \cite{2001ApJ...562.1012E}. The code evolves the orbits and spins simultaneously and accounts for mutual perturbations due to the Newtonian gravitational interactions, quadrupolar distortion of the star due to rotation and tides, tidal friction in the constant-time lag approximation, and general relativity. 

We use a Monte Carlo approach to create a suite of simulation outcomes for different system configurations. As discussed in Section \ref{sec:deepdive}, the orbital configurations resulting from the resonance interactions are dictated by $\eta_{\star}$ and $\eta_{GR}$. We thus choose key parameters to vary based on $\eta_{\star}$ and $\eta_{GR}$.  All relevant variables are defined in Table \ref{tab:random_values}, but not all of these are varied in each simulation, as we will describe in more detail in the next section. 
The viscous timescale $t_v$ is related to the tidal quality factor $Q$ of the planet by\footnote{For the stellar tidal quality factor, $k_{2,1}$ would be replaced by $k_{2\star}$, $m_1$ and $R_1$ would be replaced by $M_{\star}$ and $R_{\star}$, and $t_{v,1}$ would be replaced by $t_{v,\star}$. Our standard $t_{v,\star}$ values correspond to $Q_{\star} \approx 10^6-10^7$. }
\begin{equation}
    Q_1 = \frac{4}{3}\frac{k_{2,1}}{(1+2k_{2,1})^2} \frac{Gm_1}{R_1^3} \left(\frac{a_1^3}{GM_{\star}}\right)^{1/2} t_{v,1}.
\end{equation}

The simulations are designed to model the system's evolution just after the disk-driven resonance encounter has taken place. Given a set of parameters, we calculate the critical inclination (equation \ref{eq: I_crit}) and set this as the inner Neptune's initial inclination. To account for the fact that the initial inclination resonance occurs during the star's pre-main sequence phase while the subsequent long-term evolution is during the main sequence phase, we determine the critical inclination using $R_{\star} = 1.3 \ R_{\odot}$ and $P_{\star} =7$ days to reflect fiducial pre-main sequence values. Then, for the integration, we set the stellar parameters using the sampled variables and keep them constant. We do not evolve the stellar properties over time because we expect that tides and planet-planet interactions dominate the orbital evolution. The integrations run for 6 Gyr.

\subsection{``Tug-of-war'' simulation suites}
\label{sec: tug-of-war simulations}

\begin{figure*}
    \centering
    \includegraphics[width=\textwidth]{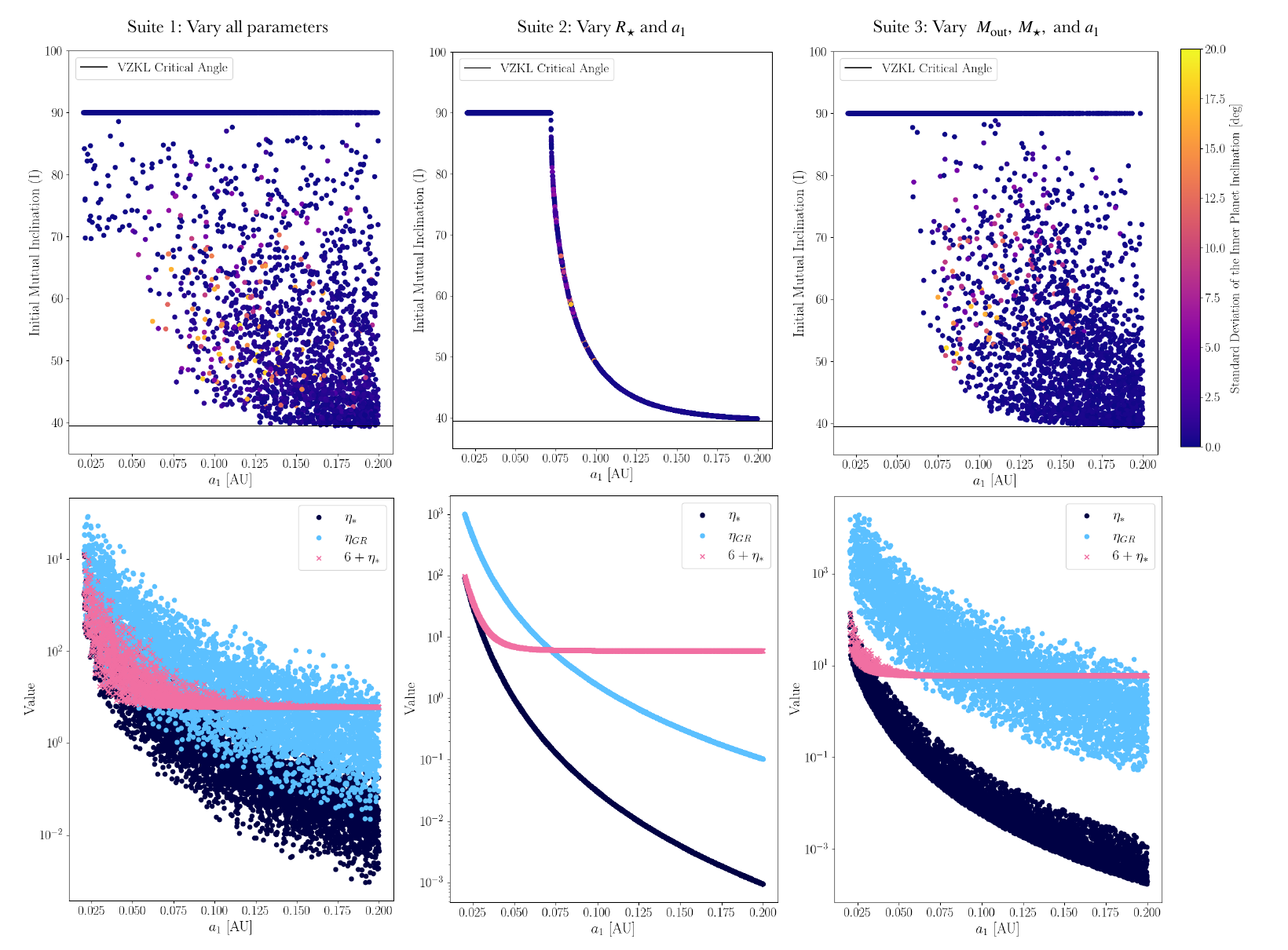}
    \caption{\textbf{Outcomes of the long-term secular simulations.} The three columns correspond to the three distinct simulation suites. In the top row, each simulation is summarized as a single dot in the space of initial inclination of the inner planet versus its semi-major axis, with the colorbar indicating the standard deviation of the inner planet's inclination throughout the simulation. In the bottom row, the initial values of $\eta_{\star}$, $\eta_{\mathrm{GR}}$ and $6+\eta_{\star}$ are plotted for each simulation to give context to the simulation results. Note that $\eta_{\star}$ and $\eta_{\mathrm{GR}}$ are computed using fiducial pre-main sequence values $R_{\star} = 1.3 \ R_{\odot}$ and $P_{\star} = 7$ days, and then these values are used to calculate the critical inclination that the inner planet is initialized with. Overall, the standard deviation is nearly zero for most runs, indicating that the inner planet's inclination generally stays very stable.}
    \label{fig:comboruns}
\end{figure*}

We define three simulation suites that span the ``tug-of-war'' between the outer Jupiter, stellar oblateness, and general relativity. In Suite 1, we sample all parameters according to the schemes outlined in Table \ref{tab:random_values}. This suite is designed to represent the full range of outcomes given any plausible parameter choices. Suites 2 and 3 are more targeted. In Suite 2, we vary $R_{\star}$ and $a_1$, and in Suite 3, we vary $M_{\star}$, $a_1$, and $m_{2}$. For all three suites, we run $\sim5,000$ different simulations. 

The results of Suites 1, 2, and 3 are shown in Figure \ref{fig:comboruns}. 
To understand the degree of variation in the inclination evolution, we compute the standard deviation of the inner planet's inclination for each run. Recall that the planet starts at the critical inclination, so a small standard deviation would indicate that the misaligned orbit is long-term stable. It is clear from Figure \ref{fig:comboruns} that very small standard deviations are the usual outcome. Specifically, for Suites 1, 2, and 3, there are respectively 375, 292, and 133 runs where the standard deviation of the inner planet's inclination is greater than $1^{\circ}$.

Suite 2 shows a distinct curve of initial mutual inclination vs. $a_1$. This is a consequence of initializing the inner planet at the critical inclination. As described earlier, Suite 2 varies $R_{\star}$ and $a_1$, but $R_{\star}$ is set to $1.3 \ R_{\odot}$ for the critical inclination calculation. Thus, there is only one degree of freedom in setting $I_{\mathrm{crit}}$, and that is reflected in the curve of the inclination versus $a_1$. For $\eta_{\mathrm{GR}} > 6 + \eta_{\star}$, $I_{\mathrm{crit}} = 90^{\circ}$ and fully polar orbits are created. The curve is also seen as an approximate lower envelope in the Suite 1 and 3 simulations where more parameters are varied.

\begin{figure}
    \centering
    \includegraphics[width=0.5\textwidth]{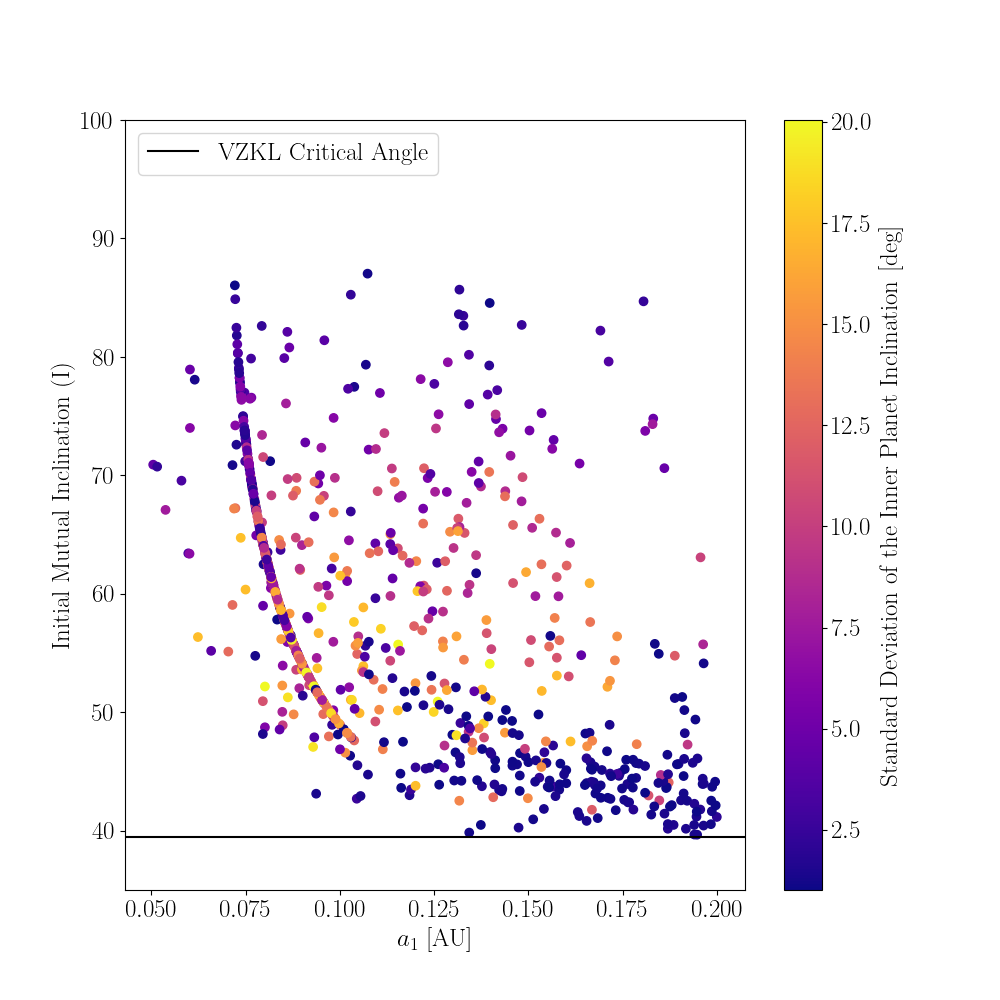}
    \includegraphics[width=0.5\textwidth]{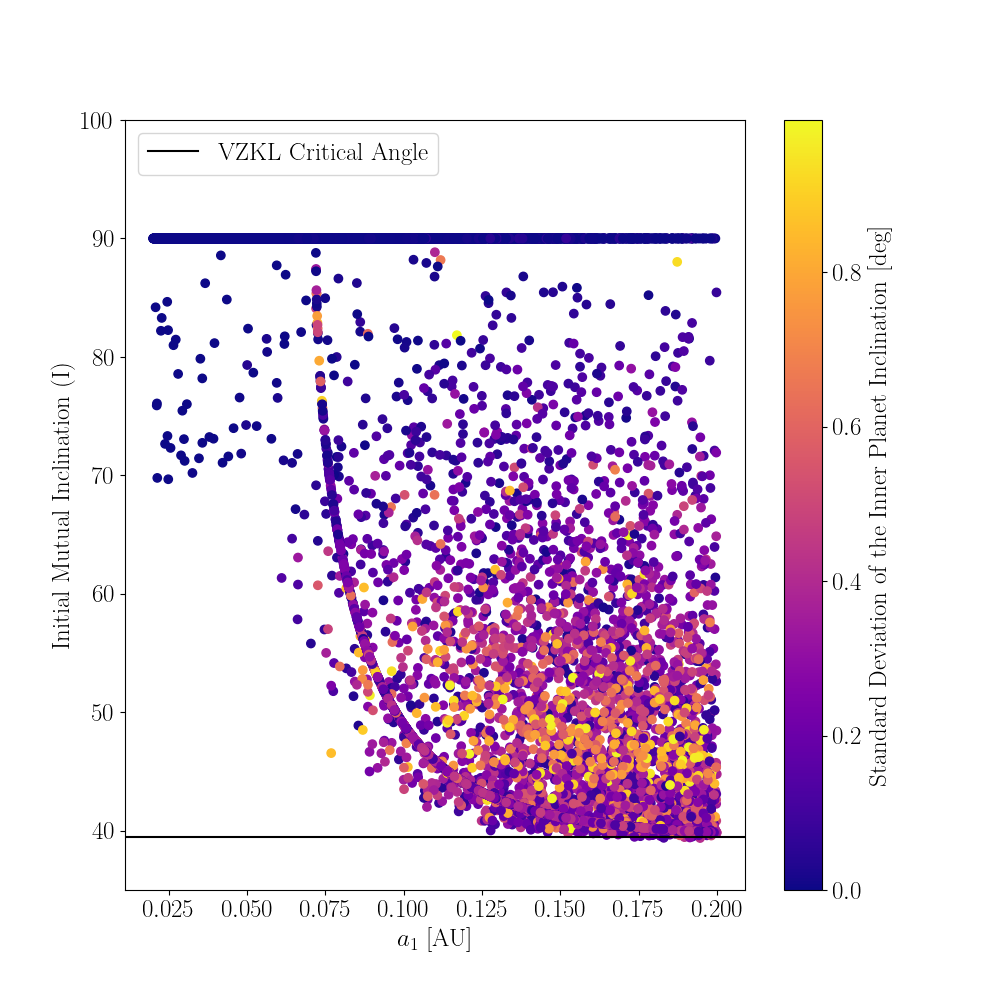}
    \caption{\textbf{Simulation outcomes split by the amount of inclination variations.} The top (bottom) panel shows the subset of simulations from all three suites where the standard deviation of the inner planet's inclination is greater (smaller) than $1^{\circ}$. Notably, all polar planets are stable and only appear in the low standard deviation subset. Even within that subset, they are the most stable.}
    \label{fig:stdsubsets}
\end{figure}

Overall, the simulation results offer a few notable takeaways. First, and most importantly for this work, the polar planets are extremely stable to tides in all three suites. While there is some deviation in the inner planet's inclination in some of the runs in each suite, the planets in polar orbits are all stable over time. This is further illustrated in Figure \ref{fig:stdsubsets}. The top (bottom) panel shows the subset of runs from all three suites with a standard deviation greater (lower) than one. The top panel includes no runs with planets in polar orbits. There are some near-polar runs that exhibit changes in inclination, but the truly polar orbits are stable. This conclusion is further cemented by the bottom panel. Even within the low standard deviation subset, the runs with the highest standard deviation are predominately concentrated around $50^{\circ}$. Our finding that polar orbits are stable to tides is consistent with the empirical results of \cite{2023A&A...674A.120A}, who compared a sample of systems where tides are expected to be weak to the full observational sample and found the prevalence of polar orbits to be roughly equivalent.

\begin{figure*}
    \centering
    \includegraphics[width=\textwidth]{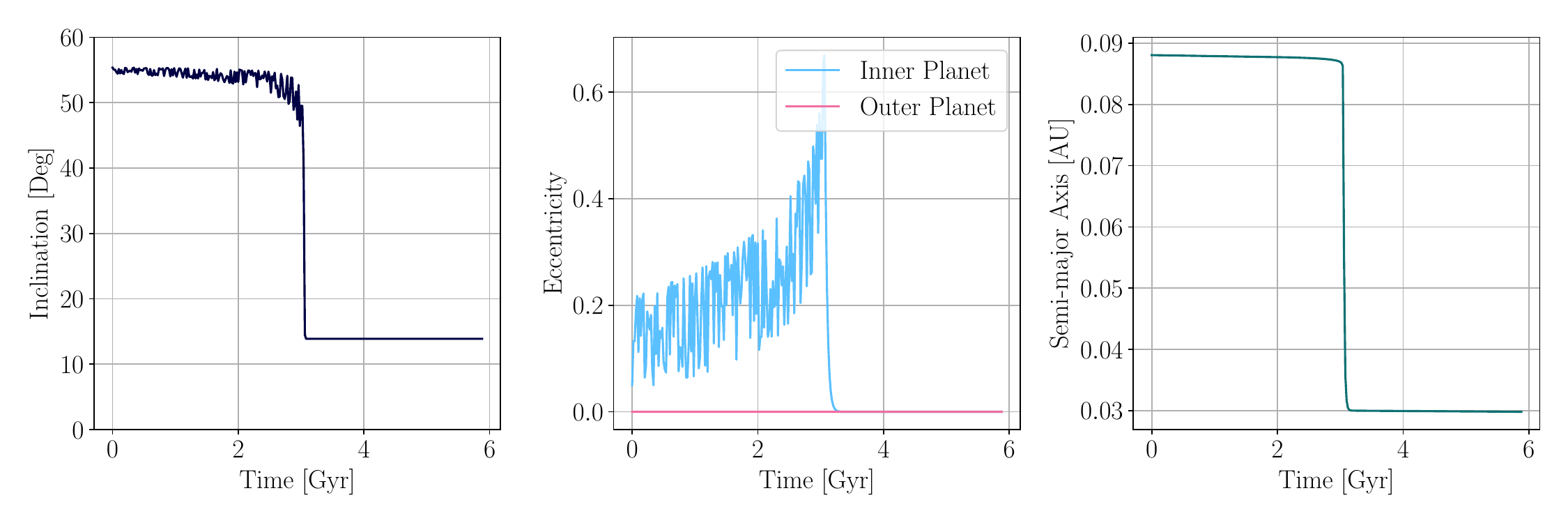}
    \caption{\textbf{Example of orbital evolution for a simulation with significant inclination variation.} We show the time evolution of the inner planet's inclination, eccentricity, and semi-major axis from left to right. This simulation is the one with the most variation in the inner planet's inclination out of all suites. The eccentricity of the inner planet rises due to interactions with the outer planet. It then undergoes rapid tidally-driven eccentricity and inclination damping and inward orbital migration.}
    \label{fig:higheststdrun}
\end{figure*}

The second key result pertains to the location of the instability hotspot for all three suites. In Figure \ref{fig:stdsubsets}, the color bar shows that the cases with the highest standard deviation of the inner planet's inclination have an initial inclination in the range $\sim45^{\circ}-80^{\circ}$. Looking more closely at these higher standard deviation runs in Figure \ref{fig:higheststdrun}, we see an evolution reminiscent of high-eccentricity migration. The eccentricity slowly increases over time, and when it gets very large, the orbit undergoes rapid tidal migration and circularization. The planet ends up in a stable circular orbit closer to the star with a lower inclination. Such outcomes are the only mechanism in which we observed large-scale inclination variations.

One caveat of these simulations is that we do not know the exact eccentricities emerging after the initial resonant encounters. As shown in Figure \ref{fig:ShortTermEvolution}, the eccentricity can be excited during the resonant encounter but then influenced by tidal evolution to stabilize at smaller values. We set the eccentricities somewhat arbitrarily but informed by the earlier simulation and the observed population of Neptune-sized planets. To test the sensitivity of this assumption, we ran a suite of simulations where the eccentricity was allowed to vary more widely and found that the stability of the polar planets as well as the location of the hotspot was unchanged. There was the same order of magnitude of runs with standard deviation of the inner planet's inclination greater than $1^{\circ}$ and the hotspot appears in the same range of initial mutual inclination. 

It is useful to contrast our results with  \cite{2013A&A...553A..39C}, who also explored the secular dynamics of similar hierarchical two-planet systems but in a slightly different parameter regime. They identified a planetary spin-facilitated mechanism that induced unexpected and significant damping of the mutual inclination in long-term orbital dynamics. To validate our own simulations and compare with their results, we run another suite of long-term simulations similar to those presented in \cite{2013A&A...553A..39C}. The set-up and results are described in Appendix \ref{sec: compare with Correia}. In short, we reproduce their findings and show that a particular regime of parameter space (namely, where the inner planet is more massive than Neptune) results in inclination damping by tens of degrees. The effect is significant, but it doesn't overlap with the parameter regime of interest in this work.

\section{Case Studies}\label{sec:casestudy}
Thus far, we have used generalized synthetic planetary systems with randomized parameters to explore the long-term dynamics following disk-driven resonance. Here we turn to observed polar Neptune systems to gain further insights. We analyze two case study planets in polar and slightly eccentric orbits: HAT-P-11 b and WASP-107 b. Both of these planets have distant giant planet companions, making them strong candidates for a disk-driven resonance origin scenario \citepalias{2020ApJ...902L...5P}. Moreover, they are actively undergoing atmospheric mass loss \citep{2020AJ....159..115K, 2024arXiv240300608G}, which could be a side effect of enhanced internal heating driven by tides from their eccentric orbits \citep{2019ApJ...886...72M}. We aim to examine the long-term evolution of the inner planets following the initial establishment of their polar orbits via disk-driven resonance. We will use dynamical simulations to track the stability of the orbits under the influence of tides. Moreover, we will explore whether the long-term orbital evolution allows us to place bounds on the planets' tidal quality factors. 

Before proceeding, we note that HAT-P-11 b and WASP-107 b were explored in two recent papers by \cite{2024arXiv240519511L} and \cite{2024arXiv240600187Y}, who showed that these planets could be explained by a high-eccentricity migration pathway induced by von Zeipel-Kozai-Lidov (ZKL) cycles. Disk-driven resonance is a potential alternative to this. We do not necessarily favor a certain formation pathway, as both theories have advantages. One advantage of disk-driven resonance is that the inner and outer planets form coplanar. Conversely, for ZKL cycles to start, the planets must first obtain a large mutual inclination, which could be produced by planet-planet scattering or a stellar flyby. However, we also note that both processes could work in concert, with disk-driven resonance producing an initial tilt and then scattering or secular interactions leading to further eccentricity and inclination changes. The outer planet in HAT-P-11 has a substantial eccentricity and possibly also a misaligned stellar obliquity, which is indicative of a history of scattering \citep{2024arXiv240519511L}.

\begin{table}[t]
\centering
\caption{\textbf{Parameters of the case study systems.}}
\label{tab:casestudy}
\begin{tabular}{|l|c|c|}
\hline
 & \textbf{HAT-P-11 b} & \textbf{WASP-107 b} \\
\hline
$M_{\star}$ ($M_{\odot}$) & 0.809 & 0.683 \\
$m_1$ ($M_{\oplus}$) & 23.39 & 30.50 \\
$m_2$ ($M_{J}$) & 1.6 & 0.36 \\
$a_1$ (AU) & Uniform(0.05, 0.06) & Uniform(0.055, 0.065) \\
$a_2$ (AU) & 4.13 & 1.823 \\
$e_1$ & Uniform(0.02, 0.95)  & Uniform(0.1, 0.9) \\
$e_2$ & 0.60 & 0.28 \\
$I$ (deg) & Critical Inclination & Critical Inclination\\
$\omega_1$ (rad) & 0.33 & 0.70 \\
$\omega_2$ (rad) & 2.50 & -2.09 \\
$\Omega_1$ (rad) & 0.00 & 0.00 \\
$\Omega_2$ (rad) & 3.14 & 3.14 \\
$P_{\mathrm{rot}}$ (days) & 29.20 & 17.10 \\
$R_{\star}$ ($R_{\odot}$) & 1.4 & 0.67 \\
$k_{2,1}$ & 0.30 & 0.30 \\
$k_{2,2}$ & 0.47 & 0.47 \\
$Q_1$ & Variable & Variable \\
$C_1$ & 0.25 & 0.25 \\
$C_2$ & 0.25 & 0.25 \\
\hline
\end{tabular}
\end{table}

\subsection{HAT-P-11 b}
The HAT-P-11 planetary system, comprising two known exoplanets, offers a compelling glimpse into the dynamical processes producing planets in slightly eccentric polar orbits. Closest to the host star is HAT-P-11 b, a Neptune-sized planet discovered utilizing the transit method \citep{2010ApJ...710.1724B}. With a mass approximately 26 times that of Earth and a radius four times larger, HAT-P-11 b orbits its host star every 4.88 days at a close distance of about 0.053 AU with a moderate eccentricity of $0.218^{+0.034}_{-0031}$ \citep{2018AJ....155..255Y}. The projected obliquity is $\lambda = 103^{+26}_{-10}$ degrees, and the 3D obliquity is $\psi = 97^{+8}_{-4}$ degrees \citep{2011ApJ...743...61S, 2021ApJ...916L...1A}. \cite{2018AJ....155..255Y} discovered the outer Jupiter-sized planet and suggested a large mutual inclination between the inner sub-Neptune and outer Jupiter, which was recently confirmed by \cite{2024arXiv240519510A}. 


Here, we explore the idea that HAT-P-11-b's polar orbit was established via a disk-driven resonance. We consider the post-disk evolution of the system and perform simulations that probe the long-term stability of HAT-P-11-b's orbit. The initial eccentricity following resonant excitation is unknown, as is the tidal $Q$ of the planet. The planet may also have undergone tidal migration, ending up in a smaller orbit than it started at. Thus, we construct a suite of 5,000 simulations with different initial eccentricities, semi-major axes, and tidal $Q$ values. Specifically, we allow the eccentricity to vary from  0.02 to 0.95, the semi-major axis from 0.05 AU to 0.06 AU, and $Q$ from $10^3$ to $10^7$, as summarized in Table \ref{tab:casestudy}. For all simulations, we start the inner planet at the critical inclination calculated using using the fiducial pre-main sequence stellar parameters ($R_{\star} = 1.3 \ R_{\odot}, \ P_{\star} = 7 \ \mathrm{days}$). The system is evolved using \texttt{KOZAIPY} for 6 Gyr.
Of the 5,000 simulations, 2,089 were excluded because the planet was engulfed by the star before reaching the end of the simulation. 

\begin{figure}
    \centering
    \includegraphics[width=0.4\textwidth]{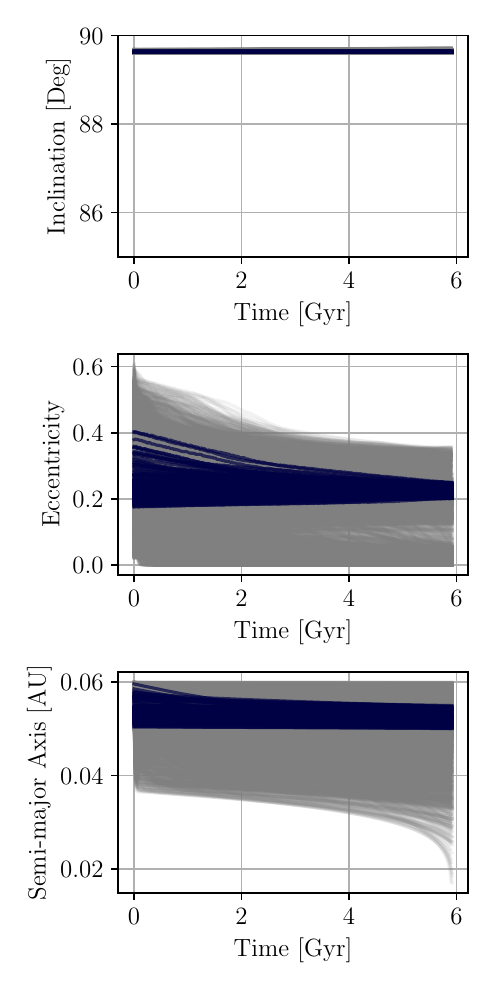}
    \caption{\textbf{HAT-P-11 b evolution.} These simulations show the long-term evolution of HAT-P-11 b immediately following a theorized disk-driven resonant encounter that established its polar orbit. From top to bottom, we show the evolution of planet b's inclination, eccentricity, and semi-major axis over time. The grey lines represent the entire sample of runs. The blue lines denote the 120 runs that end with an eccentricity and semi-major axis consistent with the observed values with a tolerance of 0.05 for eccentricity and 0.005 AU for semi-major axis. The polar orbit is extremely stable. }
    \label{fig:HATP11}
\end{figure}

The orbital evolutions of all simulations are shown in Figure \ref{fig:HATP11}. The inclination stays remarkably stable over a 6 Gyr timescale ($0.003\%$ average change in mutual inclination). The planet's eccentricity and semi-major axis evolves due to the tidal dissipation. In Figure \ref{fig:HATP11}, the blue lines show the 120 runs where the final semi-major axis and eccentricity match the observed values with tolerances of 0.005 AU and 0.05, respectively. These cases show some decrease in the semi-major axis of the inner planet. Only a certain range of $Q$ values results in an orbit that agrees with the present-day configuration. We will discuss this further in Section \ref{sec: tidal Q}.

\subsection{WASP-107 b}

\begin{figure}
    \centering
    \includegraphics[width=0.4\textwidth]{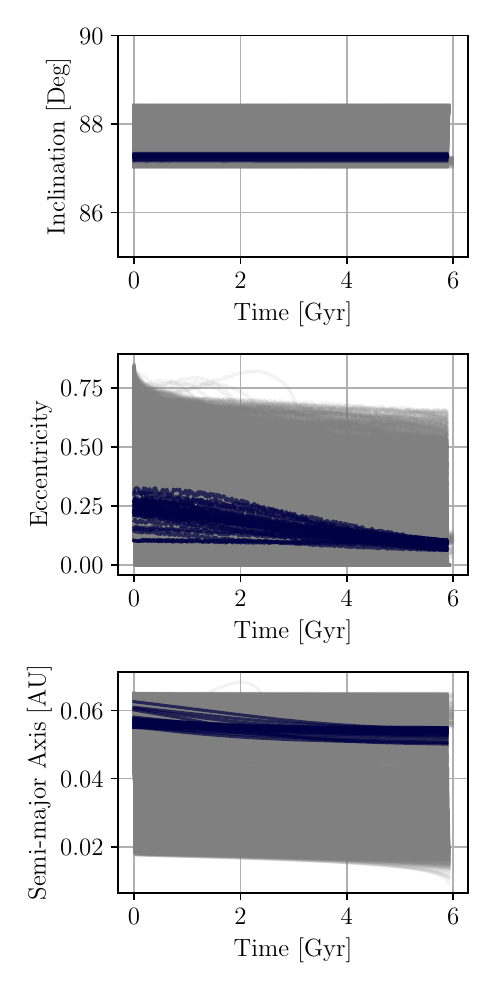}
    \caption{\textbf{WASP-107 b evolution.} Same as Figure \ref{fig:HATP11}, except for WASP-107 b.}
    \label{fig:WASP107}
\end{figure}

The WASP-107 system hosts the well-studied super-Neptune WASP-107 b \citep{2017A&A...604A.110A}. 
This planet has become a focal point for understanding the atmospheric properties and orbital dynamics of puffy planets in close proximity to their host stars. With a mass only about 30 times that of Earth yet a nearly Jupiter-sized radius, WASP-107 b completes an orbit around its host star every 5.7 days with a semi-major axis of 0.055 AU and eccentricity of $0.06\pm0.04$ \citep{2021AJ....161...70P}. The planet's orbit is polar, with a sky-projected obliquity of $\lambda = {118}^{+38}_{-19}$ degrees \citep{2017AJ....153..205D, 2021AJ....161..119R} and a 3D obliquity of $\psi = {92.6}^{+30.7}_{-1.8}$ degrees \citep{2021ApJ...916L...1A}. \cite{2021AJ....161...70P} detected a $0.36 \ M_{\mathrm{Jup}}$ companion planet on a wide eccentric orbit. Alongside dynamical studies, WASP-107 b has provided insights into the nature of inflated super-Neptunes through studies of its escaping atmosphere \citep{2020AJ....159..115K} and high internal heat flux \citep{2024arXiv240511027S, 2024arXiv240511018W}, both suggestive of strong tidal heating \citep{2019ApJ...886...72M}. The planet's unusual orbit and interior/atmosphere make it a prime subject for detailed characterization. 



Just like our case study of HAT-P-11, we run 5,000 simulations of the WASP-107 system that explore the long-term evolution of planet b after the hypothesized disk-driven resonance encounter established its polar orbit. We allow the eccentricity to vary from  0.02 to 0.95, the semi-major axis from 0.05 AU to 0.06 AU, and $Q$ from $10^3$ to $10^7$. 
Of the 5,000 simulations, 379 were excluded because the planet fell into the star before reaching the end of the 6 Gyr.  

The system evolution is shown in Figure \ref{fig:WASP107}. Like HAT-P-11 b, the inclination stays extremely stable over a 6 Gyr timescale ($0.005\%$ average change in mutual inclination). The planet's eccentricity and semi-major axis evolve due to tidal dissipation. In Figure \ref{fig:WASP107}, the blue lines show the 28 runs where the final semi-major axis and eccentricity match the observed values with tolerances of 0.005 AU and 0.05, respectively.  

\subsection{Constraining Tidal Q}
\label{sec: tidal Q}

\begin{figure}
    \centering
    \includegraphics[width=0.5\textwidth]{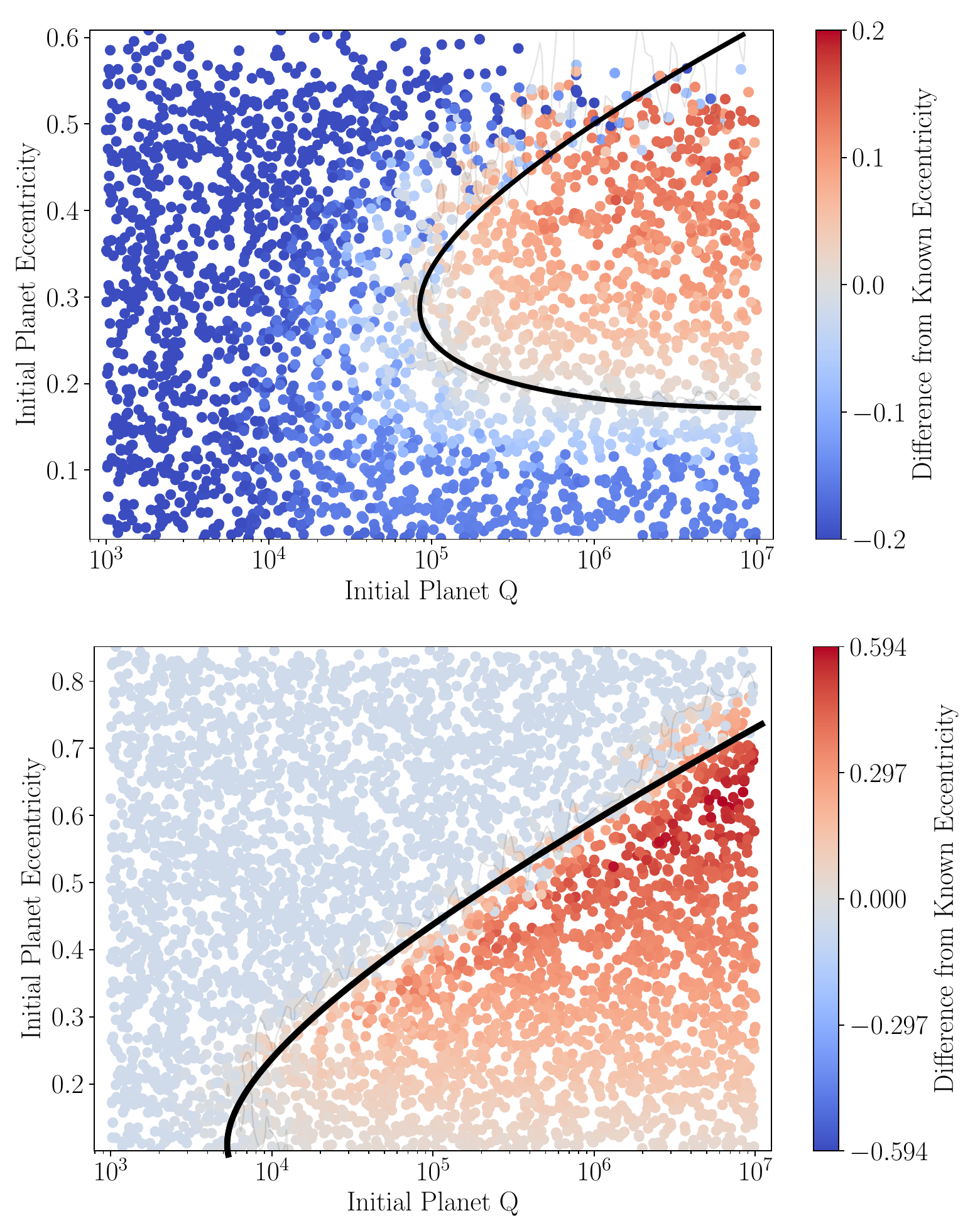}
    \caption{\textbf{Tidal $Q$ constraints.} The top and bottom panels pertain to HAT-P-11 b and WASP-107 b respectively. They show the initial eccentricity of the planet versus its initial $Q$ value. Each dot corresponds to a different simulation, and the coloring indicates the difference of the final eccentricity from the observed eccentricity, where red (blue) dots indicate a final eccentricity that is too high (too low). The black line represents an approximate contour overlaying the simulations with the closest match to the observed eccentricity. The empty corner in the upper right of the top panel is due to filtering out runs where the planet spirals into the star.}
    \label{fig:Q Values}
\end{figure}

Our simulations used a range of values for the planet's tidal quality factor $Q$, but only some of the resulting systems matched the observed system properties. Here we explore this in more detail in an effort to obtain constraints on the $Q$ values. The tidal quality factor is a metric for the efficiency of tidal energy dissipation, defined by 
\begin{equation}
     Q^{-1} = -\frac{1}{2\pi E_0}\oint \frac{dE}{dt}\,dt,
\end{equation}
where $E_0$ encapsulates the peak energy stored within the tidal distortion of the planet, and the integral represents the energy dissipated in a complete cycle \citep{1966Icar....5..375G}. Smaller values of $Q$ correspond to more efficient dissipation and vice versa. Within the Solar System, the terrestrial planets and satellites have $Q \sim 10^2$ \citep{Yoder1995, 2015JGRE..120..689W, 2016CeMDA.126..145L}, while Uranus and Neptune have $Q \sim 10^4$ \citep{1966Icar....5..375G}. 
For Neptune-like exoplanets, we expect $\sim10^4-10^5$, although the detailed values would depend on the body's specific rheological properties.

For both case study systems, we do not know the inner planet's initial eccentricity nor its $Q$, but we do have measurements for the current eccentricity. Both planets have non-circular orbits, indicating that $Q$ cannot be too low, otherwise the orbits would have circularized. We isolate the simulations that produce final eccentricities matching the observed values (blue lines in Figures \ref{fig:HATP11} and \ref{fig:WASP107}). By identifying their initial eccentricities and $Q$ values, we can obtain constraints on the $Q$ values for HAT-P-11 b and WASP-107 b within the framework of this theory. These simulations place the first $Q$ constraints for either of these planets. As shown in Figure \ref{fig:Q Values}, we find that $Q\gtrsim 10^5$ for HAT-P-11 b, and $Q\gtrsim10^4$ for WASP-107 b. The constraints depend on the initial eccentricity as well. Overall, these results suggest a level of dissipation that is relatively inefficient but consistent with expectations from the Solar System planets. Moreover, we expect that the constraints on $Q$ would further increase if we included the evolution of the planet's radius in our simulations \citep[e.g.][]{2024arXiv240600187Y, 2024arXiv240519511L}, since more significant past radius inflation would need to be balanced by a larger $Q$ to result in the same timescales of tidal evolution.

\section{Conclusions}
\label{sec:Conclusions}


The polar Neptunes are a mysterious new class of exoplanets. To investigate their origins, we extended the study of the disk-driven resonance theory proposed by \cite{2020ApJ...902L...5P}. 
This scenario suggests that the orbit tilting happens early, within the first $\sim10$ Myr of the system's lifetime. Before this work, there had not yet been investigations of the planet's subsequent $\sim$Gyrs of orbital evolution, during which tides might feasibly realign the orbits. We performed a comprehensive analysis of the impact of tidal evolution on both the initial excitation of the polar orbits and their long-term stability. Our main takeaways are as follows:
\begin{enumerate}
\item Short-term $N$-body integrations of the initial disk-driven resonance encounter reveal that tidal effects do not change the inclination evolution resulting from the resonance, but they can impact the eccentricity, leading it to stabilize at a smaller value than seen in a tides-free simulation. A broader parameter space exploration would be valuable in future work. 
\item Long-term secular simulations show that polar Neptune orbits are extremely stable. The orbits do not realign as a result of tidal dissipation in the star or in the planet. This result implies that there is no constraint on when the polar misalignment must be initially established, and thus a primordial excitation theory such as disk-driven resonance is in no tension with the observations. This result is consistent with \cite{2023A&A...674A.120A}.
\item On the other hand, systems emerging from the initial disk-driven resonance with a more moderate, non-polar obliquity in the range of $\sim45^{\circ}-80^{\circ}$ can often undergo substantial realignment due to tidal interactions. In some cases, these interactions are mediated by effects of spin-orbit coupling. 
\item HAT-P-11 and WASP-107 are exemplary case study systems, as they contain both polar Neptunes and outer giant planets. Long-term secular simulations show stable orbits for the polar Neptunes. Moreover, working within the framework of disk-driven resonance, we constrain the tidal quality factors of the planets to $Q \gtrsim 10^5$ for HAT-P-11 b and $Q \gtrsim 10^4$ for WASP-107 b, similar to Uranus and Neptune. Disk-driven resonance is not the only possible formation pathway for these planets, but it is consistent with available constraints.
\end{enumerate}

Our theoretical exploration of disk-driven resonance can be complemented with further observational tests. Disk-driven resonance predicts that the polar Neptunes are accompanied by distant giant planets with small stellar obliquities and $\sim90^{\circ}$ mutual inclinations with respect to the inner Neptunes \citepalias{2020ApJ...902L...5P}. This can be tested with future observations that constrain the full 3D architectures of systems containing polar Neptunes, including the stellar obliquities of the outer giant planets and the mutual orbital inclinations. Such constraints are already possible through the usage of Hipparcos and Gaia astrometry \citep[e.g.][]{2024arXiv240519510A}, but they will become significantly more prevalent with Gaia DR4 \citep{2023AJ....166..231E}. Moreover, in systems where outer giant planets have not yet been discovered, many more long-period perturbers may soon be identified using long-term radial velocity monitoring and Gaia astrometry \citep[e.g.][]{2014ApJ...797...14P, 2021ApJS..255....8R, 2023AJ....166..231E}. There is thus a promising future ahead for a more complete understanding of polar Neptune systems and their enigmatic origins.

\section{Acknowledgments}
This material is based upon work supported by the National Science Foundation under Grant No. 2306391. We thank the NSF for their support. We thank Cristobal Petrovich, Gu\dh mundur Stef\'ansson, Alexandre Correia, and James Owen for helpful conversations. We are grateful to the Yale Center for Research Computing for the use of the research computing infrastructure, and we especially thank Tom Langford for guidance and support using the Grace cluster. Additionally, we gratefully acknowledge access to computational resources through the MIT Engaging cluster at the Massachusetts Green High Performance Computing Center (MGHPCC) facility.

\appendix 

\vspace{-0.5cm}
\section{Secular Spin-Orbit Resonance}
\label{sec: secular spin-orbit resonance}

The early orbital evolution responsible for the secular inclination resonance encounter could potentially result in other resonances being crossed. In particular, we are interested in the secular spin-orbit resonance, which involves a commensurability between the orbit nodal precession frequency, $g = \dot{\Omega}$, and the planetary spin-axis precession frequency, $\alpha$. This resonance excites the planetary obliquity, and a non-zero planetary obliquity results in another source of tidal dissipation (in addition to eccentricity). Secular spin-orbit resonances put planets into so-called ``Cassini states'' \citep{1966AJ.....71..891C, 1969AJ.....74..483P}, equilibrium configurations of a body's spin pole in a uniformly precessing orbit frame. These have been well-studied in both Solar System and exoplanetary system contexts \citep[e.g.][]{2019NatAs...3..424M, 2019A&A...623A...4S, 2019ApJ...876..119M, 2020ApJ...903....7S, 2022MNRAS.513.3302S, 2024ApJ...961..203M}. 

To excite the planetary obliquity, the ratio of frequencies, $|g|/\alpha$, must cross unity from above. That is, a necessary condition for resonance crossing is that $|g|/\alpha > 1$ initially. Cassini states are strictly defined only for uniform orbital precession when $g = \dot\Omega$ and the inclination $I$ are constant. In cases where there are multiple sources of perturbations, including the scenario considered here, the precession is non-uniform. The planet's inclination and node evolution comprises a superposition of several modes with multiple frequencies \{$g_i$\}. However, for well-separated frequencies, the resulting spin vector equilibria are ``quasi-Cassini states,'' which behave approximately as Cassini states with $g$ equal to one of the $g_i$ modes. 

In the system considered here, a secular spin-orbit resonance for the inner planet would involve the fastest $g_i$ frequency since $\alpha$ is generally fast due to the inner planet's close-in orbit. The full set of $g_i$ frequencies can only be calculated through a complete secular analysis of the system, but they can be approximated by pairwise interactions. The fastest frequency for the inner planet is $g_{\star,1}$, the nodal precession frequency of the inner planet induced by stellar oblateness \citep{2019NatAs...3..424M}.
This frequency decreases over time due to stellar contraction and rotational spin down, meaning that $|g_{\star,1}|/\alpha_1$ evolves in the right direction for resonance crossing. We can thus assess the conditions in which the inner planet encounters the resonance by considering the initial value of $|g_{\star,1}|/\alpha_1$ 
across the parameter space. 

\begin{figure}[h]
   \centering
    \includegraphics[width=8cm]{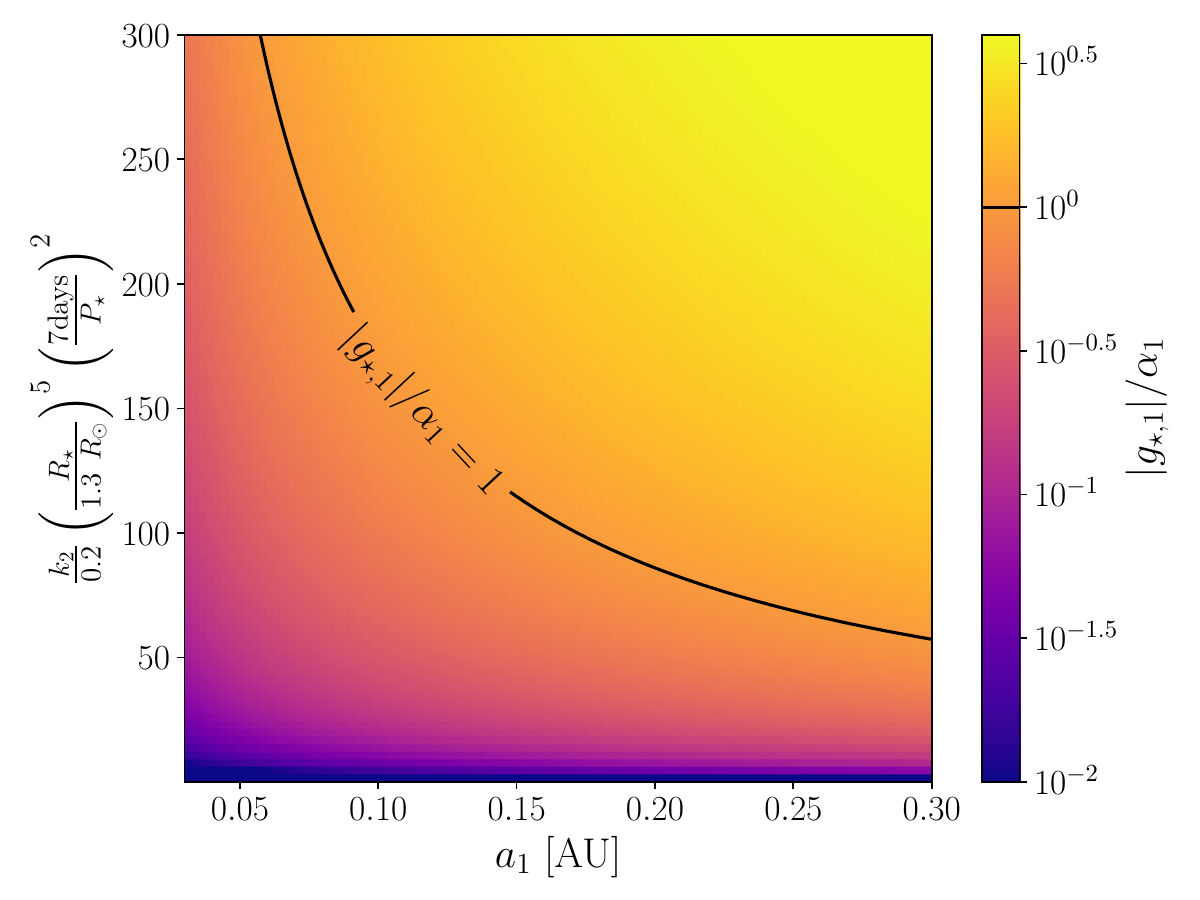}
    \caption{\textbf{Parameter space for planetary obliquity excitation.} The heatmap indicates the initial value of $|g_{\star,1}|/{\alpha_1}$ as a function of $a_1$ and a combination of stellar parameters. The planet is given fiducial parameters equal to $m_1 = 10 \ M_{\oplus}$, $R_1 = 3 \ R_{\oplus}$, $k_{2,1} =0.4$, and $C_1 = 0.25$. Obliquity excitation is expected when $|g_{\star,1}|/{\alpha_1}>1$ for initial values. However, for the parameter space that leads to polar orbits, $|g_{\star,1}|/{\alpha_1} \ll 1$.}
    \label{fig: g_over_alpha}
\end{figure}

Figure \ref{fig: g_over_alpha} shows the initial value of $|g_{\star,1}|/\alpha_1$ in a parameter space of $a_1$ on the x-axis and a combination of stellar parameters on the y-axis.
Obliquity excitation is expected when ${|g_{\star,1}|/\alpha_1>1}$ initially. The figure can be compared with \citetalias{2020ApJ...902L...5P} Figure 4, where the outcomes of the disk-driven resonance are shown in the same parameter space. We find that the region of parameter space where polar orbits are created correspond to $|g_{\star,1}|/\alpha_1 \ll 1$, far outside of the region where secular spin-orbit resonance is expected. This indicates that planets that obtain polar orbits through disk-driven resonance probably do not also obtain high obliquities, at least not through a simultaneous secular spin-orbit resonance encounter. 

\section{Comparison with \cite{2013A&A...553A..39C}} \label{sec: compare with Correia}

\cite{2013A&A...553A..39C}, hereafter \citetalias{2013A&A...553A..39C},  studied the dynamics of hierarchical two-planet systems and found that there can be a significant and surprising source of inclination damping. This does not actually result from a direct consequence of tidal effects on the orbit but rather from a complex interaction between tidal effects on the spin and secular eccentricity oscillations that lead to a gradual pumping of eccentricity and damping of inclination. Since we don't see widespread inclination damping in our simulations, it is useful to compare to \citetalias{2013A&A...553A..39C}. They considered cases where the inner planet was more massive and on a wider orbit than the close-in Neptunes we are studying here. This suggests that we do not see inclination damping simply because our systems are in a different parameter regime. We can confirm this interpretation and simultaneously validate our simulations by exploring cases with parameters similar to those used in \citetalias{2013A&A...553A..39C}. 

\begin{figure}[h]
    \centering
    \includegraphics[width=\textwidth]{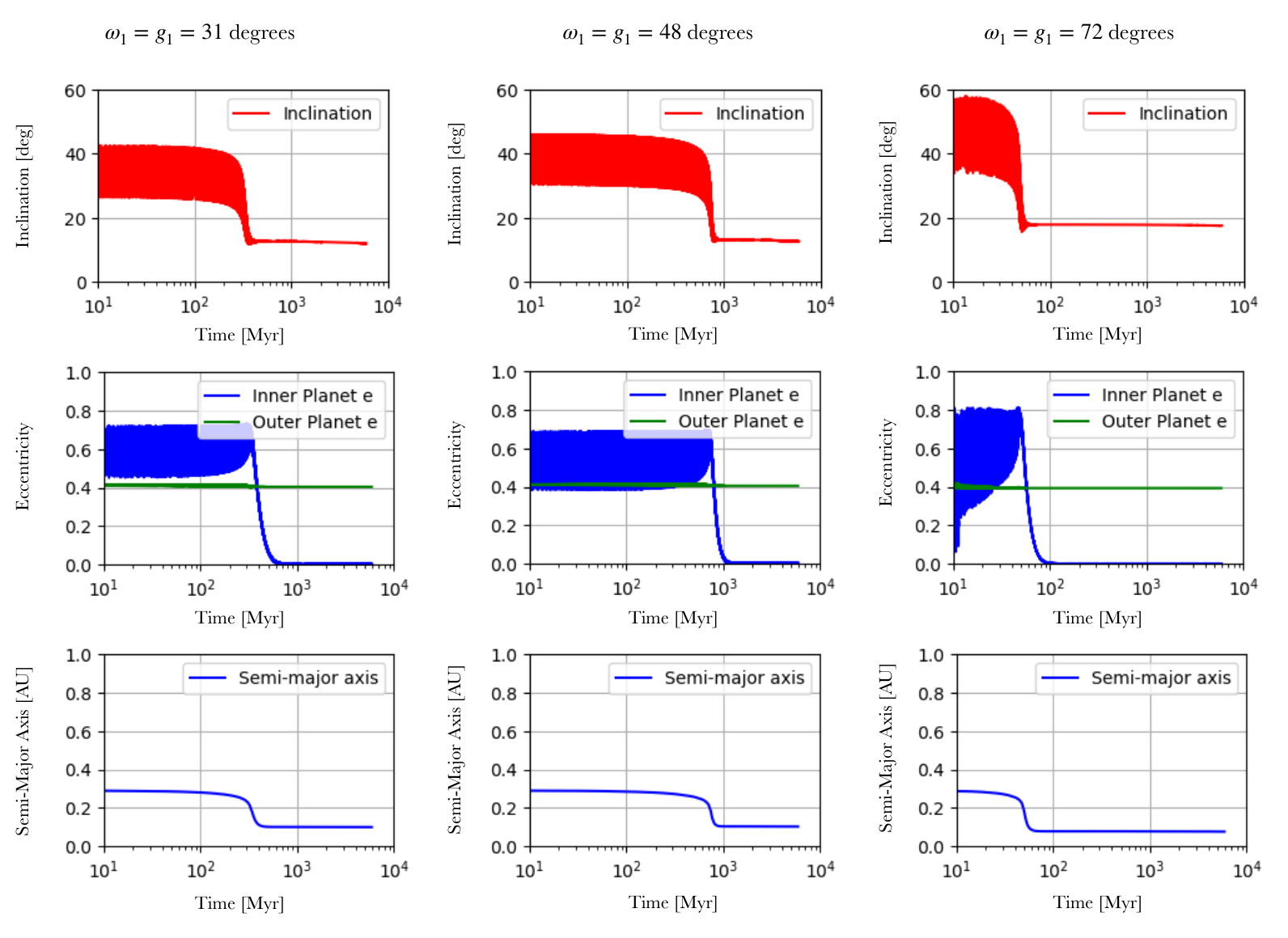}
    \caption{\textbf{Simulation validation through recreation of \citetalias{2013A&A...553A..39C} Figure 6.} Long-term evolution of the HD 74156 system with $I_0 = 40^{\circ}$ for different values of the argument of the periastron. The top, middle, and bottom panels show the mutual inclination $I$, eccentricities $e_1$ and $e_2$, and semi-major axis $a_1$, respectively. We reproduce this figure from \citetalias{2013A&A...553A..39C} to validate our simulations, which use a different numerical scheme.}
    \label{fig:Correia Test}
\end{figure}

We first reproduce a simulation performed by \citetalias{2013A&A...553A..39C} of the system HD 74156, which contains two giant planets on eccentric orbits \citep{2004A&A...414..351N, 2015ApJ...800...22F}. We use the same set-up as \citetalias{2013A&A...553A..39C} Figure 6, and the results are shown in Figure \ref{fig:Correia Test}. The orbital evolution is nearly identical to that seen in \citetalias{2013A&A...553A..39C}. The inner planet experiences significant semi-major axis, inclination, eccentricity damping as a result of the coupled spin-orbit and tidal evolution. The close agreement of our simulation results with \citetalias{2013A&A...553A..39C} offers a strong validation of our numerical method. 

We further explore a comparison to \citetalias{2013A&A...553A..39C}'s findings using a larger set of simulations. Similar to Section \ref{sec: tug-of-war simulations}, we create a simulation suite of the long-term orbital evolution of synthetic systems where the inner planet is initialized at the critical inclination. However, we use parameters that are more similar to the systems explored in \citetalias{2013A&A...553A..39C}. The parameters that differ include the inner planet's semi-major axis and the planet masses. The outcomes are consistent with \citetalias{2013A&A...553A..39C} in that they show significant inclination damping. About 89\% of the runs resulted in a standard deviation of the inner planet's inclination greater than $1^{\circ}$, compared to only 6\% of our earlier runs in the disk-driven resonance regime that had this high level of inclination variation (e.g. Figure \ref{fig:stdsubsets}). Clearly, these systems are in a different regime than the polar Neptunes we explored in Section \ref{sec: tug-of-war simulations}. 

\bibliographystyle{aasjournal}
\bibliography{main}

\end{document}